\documentclass[conference]{IEEEtran}
\IEEEoverridecommandlockouts
\usepackage{cite}
\usepackage{amsmath,amssymb,amsfonts}
\usepackage{amsmath,amssymb,dsfont,color,float,empheq}
\usepackage{algorithmic}
\usepackage{graphicx}
\usepackage{textcomp}
\usepackage{xcolor}
\usepackage{tikz}
\usepackage{pgfplots}
\usepackage{pgfplotstable}
\usepackage{tikz,pgfplots}
\usepackage[utf8]{inputenc}
\usepackage{subfigure}
\usepackage{graphicx,wrapfig,multicol,multirow}
\usepackage{psfrag,wrapfig,subfig}
\usetikzlibrary{positioning,shadows,shapes,backgrounds,calc}
\usetikzlibrary{positioning}
\usepackage{makecell}
\usepackage{multirow}
\usepackage{array}
\usepackage{caption}
\usepackage{balance}

\def\BibTeX{{\rm B\kern-.05em{\sc i\kern-.025em b}\kern-.08em
    T\kern-.1667em\lower.7ex\hbox{E}\kern-.125emX}}

\pgfplotsset{compat=1.18}

\begin{document}

\title{
 Multidimensional Voronoi Constellations vs. Short Blocklength Probabilistic Shaping:  
A Comparison for Multilevel Coding Approach
}

\author{\IEEEauthorblockN{Yajie~Sheng$^1$, Bin~Chen$^{1,*}$, Yi Lei$^1$, Jingxin~Deng$^1$, Jiwei~Xu$^1$,  Mengfan Fu$^2$, Qunbi Zhuge$^2$, Shen Li$^3$}

\thanks{This work is supported by the NSFC Programs (62171175 and 62001151), the Fundamental Research Funds for the Central
Universities under Grant JZ2024HGTG0312, and State Key Laboratory of Advanced Optical Communication Systems and Networks, Shanghai Jiao Tong University, China (2023GZKF017).}

\IEEEauthorblockA{$^1$ Department of Computer Science and Information Engineering, Hefei University of Technology, Hefei, China\\
$^2$ State Key Laboratory of Advanced Optical Communication Systems and Networks, Department of Electronic Engineering, \\Shanghai Jiao Tong University, Shanghai, China\\
$^3$ Centre of Optics, Photonics and Lasers (COPL), Department of Electrical and Computer Engineering, \\Université Laval, Québec, Canada\\
$^*$Corresponding author: bin.chen@hfut.edu.cn}
}

\maketitle

\begin{abstract}
Performance of concatenated multilevel coding 
with probabilistic shaping (PS) and Voronoi constellations (VCs) is  analysed over AWGN channel.
Numerical results show that VCs provide up to 1.3~dB 
SNR gains over 
PS-QAM with CCDM  blocklength of 200.

\end{abstract}


\section{Introduction}
The ever-increasing demand for high transmission rates to support the Internet’s exponential traffic growth is pushing the study of forward error correction (FEC) schemes with higher-order modulation, a combination known as coded modulation (CM). 
Recently, multilevel coding (MLC)~\cite{Wachsmann1999} with multi-stage decoding (MSD) has been considered as an effective way to substantially reduce the power consumption of soft-decision (SD) FEC decoding for higher order modulation formats \cite{Bisplinghoff2017}. Unlike conventional bit-interleaved coded modulation (BICM) that protects all the bit levels equally, MLC only encodes the least reliable bits (LRBs) by an inner FEC code.

For additive white Gaussian noise (AWGN) channel, conventional high-order quadrature amplitude modulation (QAM) formats with uniform distribution have 1.53~dB gap \cite{Forney1984} to Shannon capacity. To overcome the capacity constraints, constellation shaping methods have received considerable attention, which consist of two categories: probabilistic shaping~\cite{Fehenberger2016} and geometric shaping (GS)~\cite{ChenBin2023JLT}. Hybrid probabilistic and geometric shaping has been also investigated
as in~\cite{Soleimanzade2023}.

Probabilistic shaping, which changes the occurrence probability of constellation points by using distribution matching (DM), has been extensively studied because of its shaping gain and rate adaptivity for systems with fixed FEC~\cite{Böcherer2015}. Implementations of PS are realized via various approaches, such as constant composition distribution matchers (CCDM)~\cite{Schulte2016} and enumerative sphere shaping (ESS)~\cite{Amari2019}.  Particularly, concatenated MLC with PS~\cite{Yoshida2020,Sugitani2021,Matsumine2022} has attracted great attention in recent years. The previous results demonstrate that the integration of PS with a MLC system could achieve better performance-complexity trade-offs with much lower complexity.

Geometric shaping is also a popular technique to  enhance the achievable rate by adjusting the locations of constellation points. Voronoi constellations are considered as an effective geometric shaping method which inherently performs a joint shaping of multiple dimensions, yielding a good trade-off between shaping gain and complexity. Multidimensional (MD) VCs have demonstrated better bit error rate (BER) performance compared to traditional Gray-labeled QAM formats \cite{Mirani2021,Li2021ISIT}. Concatenated MLC with MD VCs can further achieve higher signal-to-noise ratio (SNR) gains with a markedly low complexity over BICM \cite{li2023arXiv}, and have been experimentally demonstrated over single-mode fiber and multi-core fiber transmission \cite{HeZonglong2024,Zhao:24}.
 
In this paper, we compare the performance of concatenated MLC  with probabilistic shaping and Voronoi constellations for AWGN channel. To the best of our knowledge,
this is the first time such comparison is made between PS and VC formats.
Specifically, we apply two different distribution matchers for two blocklengths.
Numerical results show that VCs can
achieve up to 0.61~dB and 1.3~dB SNR gains over PS-64QAM and PS-256QAM with CCDM blocklength of 200, respectively. Moreover, it also demonstrates that high-order PS-256QAM with lower shaping rate performs slightly better than VCs, which comes at the cost of high probabilistic shaping complexity.

\begin{figure*}[!tb]
    \centering   
    \includegraphics[width=52em]{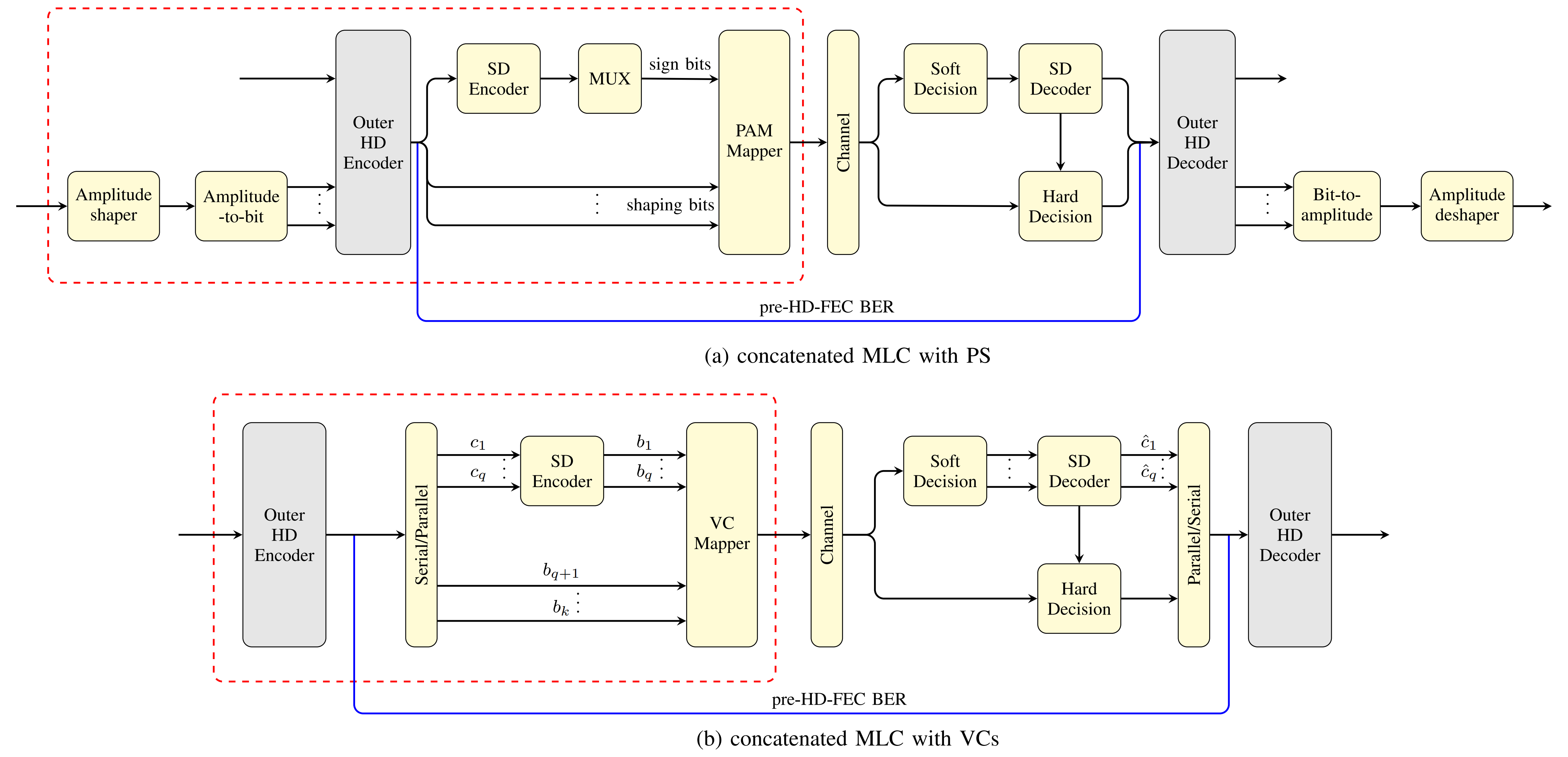}
    \caption{Block diagram of concatenated MLC with PS and VCs.}
    \label{fig:PS-MLC}
\end{figure*}

\section{Multilevel coding with PS and VCs}
In this section, we introduce the general coded modulation scheme  of PS and VCs concatenated with MLC, respectively, as shown in Fig.~\ref{fig:PS-MLC}(a) and (b).

\subsection{Probabilistic Shaping with MLC}

Probabilistic shaping is based on identical $2^{m} $-ary pulse amplitude modulation (PAM) constellations $\mathcal{X} =\left \{ \pm 1,\pm 3, \cdots ,\pm 2^{m} -1 \right \} $, which can also be 
factorized as $\mathcal{X} =\mathcal{A} \times \mathcal{S} $, where $\mathcal{A} =\left \{ 1,2,\cdots ,\left ( 2^{m}-1  \right )\right \} $ and $\mathcal{S} =\left \{ -1, +1\right \} $. QAM constellations are  Cartesian product of PAM symbols by assigning two PAM symbols to real and imaginary components.

The block diagram of concatenated MLC with PS is illustrated in Fig.~\ref{fig:PS-MLC} (a). At the transmitter, an uniform bit sequence is transmitted to the amplitude shaper that is a critical technique for implementing shaping and deshaping of shaped sequences probabilistically. The rate of this shaper (bits/1D-sym) is defined as shaping rate $R_{s} = L/N$ (ratio of the shaper's input bits $L$ and output shaped amplitudes $N$).
In this work, CCDM~\cite{Schulte2016}, which is based on $sequential$ arithmetic coding, and ESS~\cite{Amari2019}, which considers amplitude sequences satisfying a sphere of maximum-energy, are considered as approaches to realizing this conversion. Next, the amplitude symbols from the amplitude shaper are mapped into nonuniform bits as the most reliable bits (MRBs). An amplitude symbol consists of $(m - 1)$ shaping bits.
After all the shaping bits and another uniform bit sequence as LRBs are encoded by HD-FEC, the inner SD encoder is performed on the LRBs resulting in sign bits and the shaping bits are directly fed into the PAM symbol mapper. The outer HD-FEC is assumed to be systematic to maintain the distribution of amplitude symbols.
Without considering the code rate of outer HD-FEC, the overall information rate of concatenated MLC with PS is then given by
\begin{equation}
    R_{t} =2\times (R_{s} +R_{\mathrm{inner} } ),
\end{equation}
where $R_{\mathrm{inner}}$ denotes the inner SD-FEC code rate.

Bit-labeling of PAM symbol considered is based on the labeling strategy in \cite{Matsumine2022}, which not only is compatible with PS that labeling should be sign-bit decomposable but also meets the requirement of MLC that the minimum Euclidean distance typically increases after decoding the LRBs in the multi-stage decoding.
At the receiver side, the log likelihood ratio (LLR) is calculated by soft decision based on the mapping of received symbols and the probability distribution of the constellations. After assembling the output of the SD-FEC decoder, hard decision (HD) is performed, followed by an amplitude deshaper to retrieve original information
bits. Outer HD encoder and decoder are adopted to achieve a target BER of the specific system, which ensures the strict requirement for high reliability in a communication system.

\subsection{Voronoi Constellation with MLC}

VCs are structured multidimensional lattice-based constellations comprising a cubic coding lattice and and a shaping lattice.
The general definition of VCs is based on an arbitrary lattice partition $\Lambda /\Lambda _{s} $ that is defined by Forney in~\cite{Forney1989} as 
\begin{equation}
\Gamma \triangleq(\Lambda-\boldsymbol{a}) \cap \Omega\left(\Lambda_{\mathrm{s}}\right)
\label{VCs-definition}
\end{equation}
where $a$ is the offset vector avoiding VC points falling on the boundary of $\Omega\left(\Lambda_{\mathrm{s}}\right)$.
The coding lattice $\Lambda$ determines
the arrangement of constellation points yielding a coding gain, and the shaping lattice $\Lambda _{s} $ determines the shape of the constellation boundary leading to a shaping gain.
In this paper, we consider 24-dimensional (24D) VCs~\cite{li2023arXiv,HeZonglong2024}: $\Lambda _{24}^{72} $ (SE = 6 bits/2D-sym) and $\Lambda _{24}^{96} $ (SE = 8 bits/2D-sym). A generator matrix of the 24D Leech lattice $\Lambda _{24} $ is given in~\cite{Conway}.

The block diagram of concatenated MLC with VCs is depicted in Fig.~\ref{fig:PS-MLC} (b).
Information bits are first input to the outer HD encoder whose output is partitioned into $q$ parallel streams $c_{i},\cdots,c_{q}$ and $k-q$ parallel streams $b_{q+1},\cdots,b_{k}$, where $k$ denotes the number of bits that each VC symbol carries. Then $c_{i}$ ($i=1,\cdots,q$) as LRBs are encoded by the inner encoder to $b_{i}$, whereas $b_{q+1},\cdots,b_{k}$ as MRBs are directly fed into the VC mapper. VC mapping can be divided into two stages, i.e, initially converting from binary labels to integers followed by mapping these integers to VC points. The overall information rate per 2D-sym (bits/2D-sym) of concatenated MLC with $n$-dimensional VCs is defined as 
\begin{equation}
    R_{\text {t }}=\frac{2\left((k-q)+q R_{\text {inner}}\right)}{n}.
\end{equation}

After passing through channel, similar to PS-MLC, MSD is also employed where the inner SD decoder is applied to decode the LRBs to obtain the estimated bits $\hat{c}_{i}$ \cite[Eq.~(34)]{li2023arXiv} and the remaining bits are estimated based on the estimates of the LRBs by hard decision \cite[Eq.~(37)]{li2023arXiv}. Subsequently, the estimated MRBs are then passed into the outer HD decoder along with the estimation of the LRBs after going through parallel/serial conversion.

\section{Simulation Parameters and Numerical Results}

\begin{table}[!b]
    \centering
      \caption{FEC and Modulation Parameters Used in the Numerical Simulation}
    \scalebox{1}
    {\hspace{-1.5em}

\setlength{\extrarowheight}{3.3pt}
\setlength{\tabcolsep}{2.5pt}

\begin{footnotesize}
\begin{tabular}{cccccc}
\hline\hline
 \makecell[c]{Modulation \\ Format} & \makecell[c]{CM\\Scheme} & \makecell[c]{$R_{\mathrm{inner} } $}  & \makecell[c]{$R_{s}$  \\ (bits/1D-sym)} & \makecell[c]{SD-FEC \\coded/uncoded \\bits}& \makecell[c]{$R_{t}$ \\ (bits/2D-sym)}\\

 
\hline
$\Lambda _{24}^{72} $ & MLC & $2/3$ & - & 24/48 & 5.33\\
64QAM & BICM& $8/9$  & - & 6/0 & 5.33\\ 
PS-64QAM & MLC & $4/5$  & 1.87 & 2/4 & 5.34\\ 

$\Lambda _{24}^{96} $ & MLC & $3/5$ & - & 24/72 & 7.2\\
256QAM & BICM & $9/10$ & - & 8/0 & 7.2\\ 
PS-256QAM & MLC & $4/5$ & 2.8 & 2/6 & 7.2\\ 
PS-256QAM & MLC  & $4/5$ & 1.87 & 2/6 & 5.34\\ 





\hline
\hline
\end{tabular}
\end{footnotesize}
}  
    \label{tab:my_tab}
\end{table}

\begin{figure}[!tb]
     \centering
       \hspace{-1.8em}
       \includegraphics[width=26.5em]{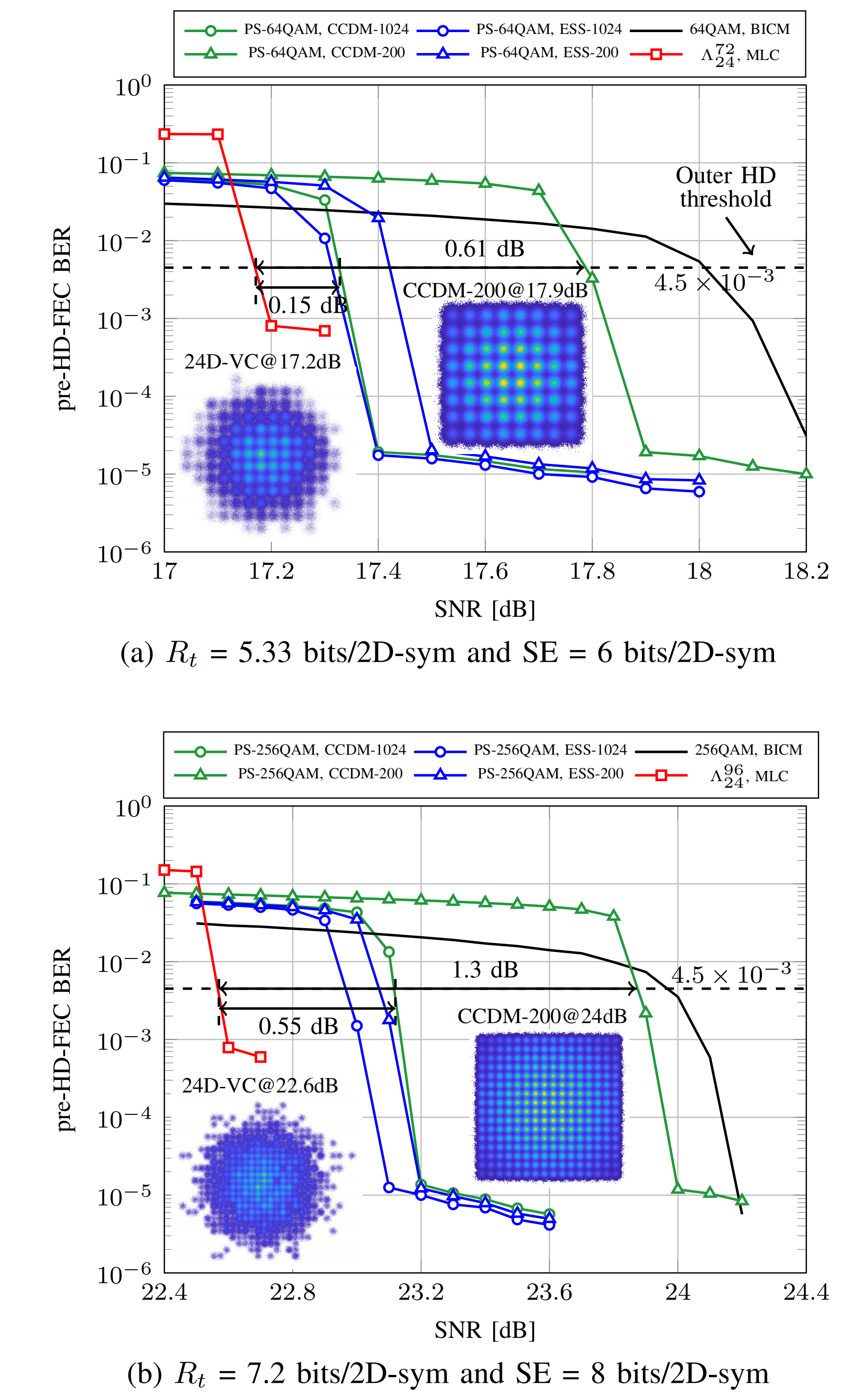}
    \vspace{0.5em}
     \caption{pre-HD-FEC BER performance of concatenated MLC with VCs and PS over AWGN channel. Uniform QAM with BICM is also shown as baseline. (a) Insets: 2D projections of VCs and PS-64QAM with CCDM for SNR = 17.2 and 17.9~dB, respectively. (b) Insets: 2D projections of VCs and PS-256QAM with CCDM for SNR = 22.6 and 24~dB, respectively.}
      \label{fig:BER1} 
            \vspace{-1em}
\end{figure}

We compare the BER performance before
HD decoder, which is called as pre-HD-FEC BER, of concatenated MLC with PS and VCs for AWGN channel. 
A set of low-density
parity-check (LDPC) codes of length 64800~bits from the digital video broadcasting-satellite-second generation (DVB-S2) standard is used for the inner SD-FEC.
The considered SD-FEC code rates and modulation parameters are listed in Table~\ref{tab:my_tab}.
MLC exhibits an error floor due to the inner uncoded MRBs, which can be mitigated by an outer HD-FEC.  
Here we assume the staircase code with a coding rate of 239/255 = 0.937 as the outer HD-FEC, which has a corresponding BER threshold of $4.5\times10^{-3}$ \cite{smith2011staircase}.

Fig.~\ref{fig:BER1} (a) shows the pre-HD-FEC BER performance between PS-64QAM and VCs concatenated MLC with SE of 6~bits/2D-sym around the information rate $R_{t}$ = 5.33~bits/2D-sym, where insets are 2D projections of VCs and PS-64QAM
for SNR of 17.2 and 17.9~dB, respectively. The pre-HD-FEC BER of 64QAM concatenated BICM is also shown as the baseline. As can be seen, VCs provide 0.15~dB gains at the BER threshold of $4.5\times 10^{-3} $ in comparison with PS-64QAM for a long CCDM blocklength of 1024. As the CCDM blocklength decreasing, performance gains of VCs are increased. For a shorter blocklength of 200, VCs outperform markedly over PS-64QAM by 0.61~dB.

Fig.~\ref{fig:BER1} (b) shows the pre-HD-FEC BER performance of PS-256QAM and VCs concatenated MLC with SE of 8~bits/2D-sym at the information rate of 7.2~bits/2D-sym. 2D projections of VCs for SNR = 22.6~dB and PS-256QAM for SNR = 24~dB are illustrated as insets. The  BER performance of 256QAM concatenated BICM is also shown as baseline. It can be observed that VCs can achieve up to 0.55~dB and 1.3~dB gains over PS-64QAM and PS-256QAM with CCDM, respectively.
The reason for larger performance gains as CCDM blocklength decreasing is that CCDM suffers from a significant rate loss at long blocklengths.

For ESS cases in Fig.~\ref{fig:BER1}, though performance gains provided by VCs over PS-QAM are decreased, superior performance of VCs can also be observed. 
In general, BER performances with CCDM or ESS are both inferior to that of VCs since VCs as advanced multidimensional modulation formats have inherently higher asymptotic shaping gain. 

Although VCs are uniform in multidimensional space, we can also observe that in insets of Fig.~\ref{fig:BER1} (a) and Fig.~\ref{fig:BER1} (b),   many  VCs points  in 2D projection are located in the low-energy regime of thereby approximating a 2D Gaussian probability distribution and a Gaussian-shaped signal, which in principle provides larger shaping gains than PS-QAM.

\begin{figure}[!tb]
    \centering
      \centering
      \hspace{-1.8em}
        \includegraphics[width=26.5em]{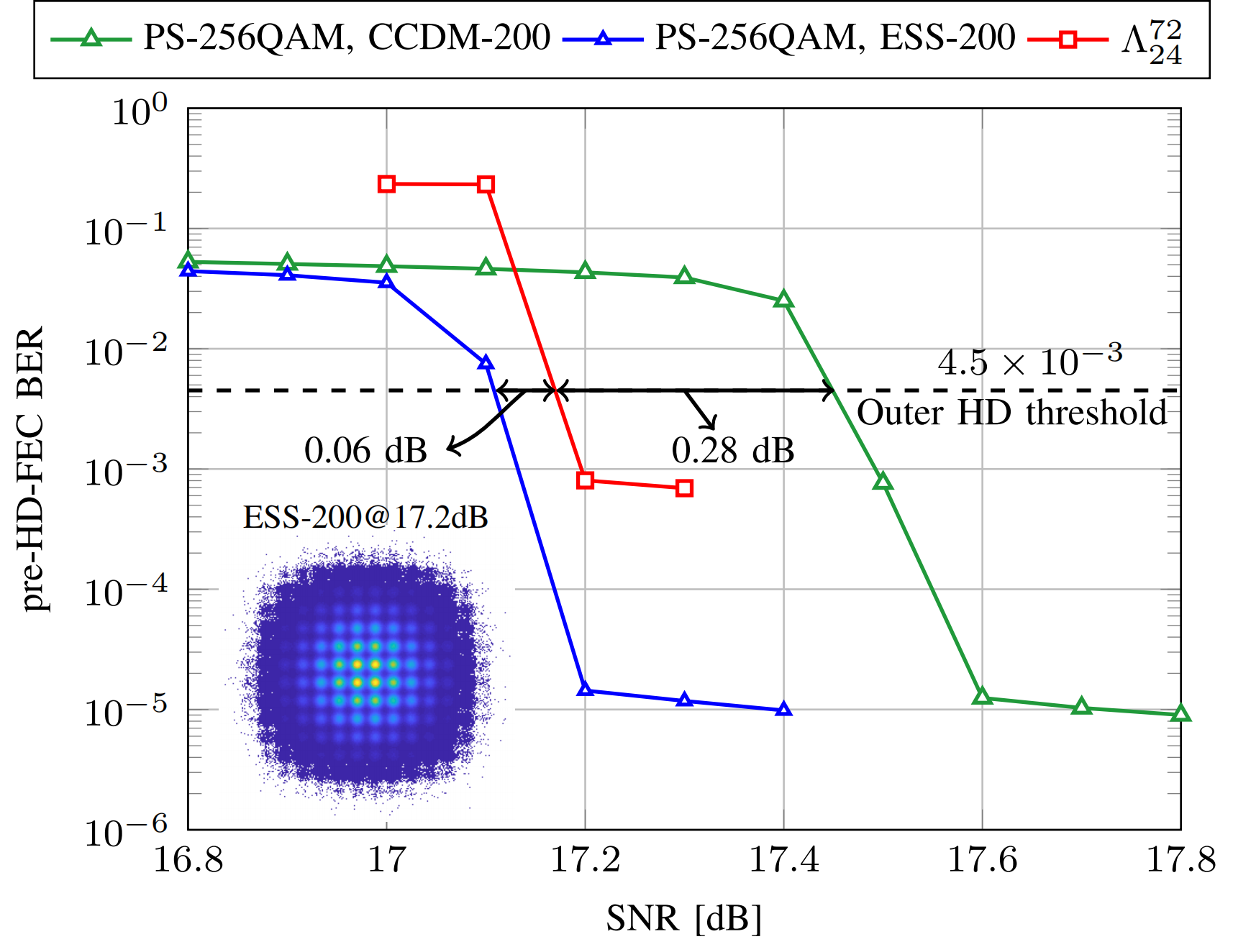}      
         \caption{pre-HD-FEC BER performance of VCs (SE = 6~bits/2D-sym) and PS-256QAM (SE = 8~bits/2D-sym) concatenated MLC at the information rate of 5.33~bits/2D-symbol. 
         The blocklength of PS-256QAM is 200. Inset:  2D projection of PS-256QAM with ESS for SNR = 17.2~dB.}
      \label{fig:256QAM-MLC-BER performance-2} 
\end{figure}

Performance losses of concatenated MLC with PS in Fig.~\ref{fig:BER1} could be interpreted as high shaping rate due to the suboptimal shaping of the system. 
Thus, 
we further compare concatenated MLC with PS-256QAM to VCs with SE = 6~bits/2D-sym around the information rate $R_{t}$ = 5.33~bits/2D-sym by decreasing the shaping rate of PS-256QAM, as illustrated in
Fig.~\ref{fig:256QAM-MLC-BER performance-2}.

For a short blocklength of 200, VCs yet give approximately 0.28~dB SNR gains over PS-256QAM with CCDM, although PS-256QAM with ESS performs slightly better than VCs, only by 0.06~dB, which simultaneously stems from both coding and shaping gains of concatenated MLC with PS. 
The 2D projection of 256QAM with ESS for SNR = 17.2~dB is also shown as inset of Fig.~\ref{fig:256QAM-MLC-BER performance-2}. 

It is worth observing that lowering shaping rate to achieve a better performance leads to a significantly higher complexity of probabilistic shaping, which makes implementation of concatenated MLC with high-order PS quite challenging in practical communication system.

\section{Conclusion}
In this paper, we demonstrated the comparison of probabilistic
shaping and Voronoi constellation formats for AWGN channel. This is the first comparison between PS and VCs, showing excellent performance of VCs over PS for short blocklengths, which makes VCs a promising candidate to be implemented in practical optics fiber communication systems.  
Moreover, despite slightly better performance of high-order PS-QAM by decreasing shaping rate, it leads to significant complexity of probabilistic shaping. 
However, excellent performance of VCs and rate adaptability of PS indicate a potential of hybrid PS and VCs concatenated MLC for future optical and wireless communications.

\balance
\bibliographystyle{IEEEtran}
\bibliography{references}

\begin{thebibliography}{10}
\providecommand{\url}[1]{#1}
\csname url@samestyle\endcsname
\providecommand{\newblock}{\relax}
\providecommand{\bibinfo}[2]{#2}
\providecommand{\BIBentrySTDinterwordspacing}{\spaceskip=0pt\relax}
\providecommand{\BIBentryALTinterwordstretchfactor}{4}
\providecommand{\BIBentryALTinterwordspacing}{\spaceskip=\fontdimen2\font plus
\BIBentryALTinterwordstretchfactor\fontdimen3\font minus \fontdimen4\font\relax}
\providecommand{\BIBforeignlanguage}[2]{{%
\expandafter\ifx\csname l@#1\endcsname\relax
\typeout{** WARNING: IEEEtran.bst: No hyphenation pattern has been}%
\typeout{** loaded for the language `#1'. Using the pattern for}%
\typeout{** the default language instead.}%
\else
\language=\csname l@#1\endcsname
\fi
#2}}
\providecommand{\BIBdecl}{\relax}
\BIBdecl

\bibitem{Wachsmann1999}
U.~Wachsmann, R.~Fischer, and J.~Huber, ``Multilevel codes: theoretical concepts and practical design rules,'' \emph{IEEE Transactions on Information Theory}, vol.~45, no.~5, pp. 1361--1391, 1999.

\bibitem{Bisplinghoff2017}
A.~Bisplinghoff, S.~Langenbach, and T.~Kupfer, ``Low-power, phase-slip tolerant, multilevel coding for {M-QAM},'' \emph{Journal of Lightwave Technology}, vol.~35, no.~4, pp. 1006--1014, 2017.

\bibitem{Forney1984}
G.~Forney, R.~Gallager, G.~Lang, F.~Longstaff, and S.~Qureshi, ``Efficient modulation for band-limited channels,'' \emph{IEEE Journal on Selected Areas in Communications}, vol.~2, no.~5, pp. 632--647, 1984.

\bibitem{Fehenberger2016}
T.~Fehenberger, A.~Alvarado, G.~Böcherer, and N.~Hanik, ``On probabilistic shaping of quadrature amplitude modulation for the nonlinear fiber channel,'' \emph{Journal of Lightwave Technology}, vol.~34, no.~21, pp. 5063--5073, 2016.

\bibitem{ChenBin2023JLT}
B.~Chen, Y.~Lei, G.~Liga, Z.~Liang, W.~Ling, X.~Xue, and A.~Alvarado, ``Geometrically-shaped multi-dimensional modulation formats in coherent optical transmission systems,'' \emph{Journal of Lightwave Technology}, vol.~41, no.~3, pp. 897--910, 2023.

\bibitem{Soleimanzade2023}
A.~Soleimanzade, M.~Amin~Soleimanzade, A.~Abolfathimomtaz, M.~Ardakani, and H.~Ebrahimzad, ``Hybrid probabilistic-geometric shaped constellations to combat fiber non-linearity,'' in \emph{IEEE International Conference on Communications (ICC)}, 2023, pp. 1946--1951.

\bibitem{Böcherer2015}
G.~Böcherer, F.~Steiner, and P.~Schulte, ``Bandwidth efficient and rate-matched low-density parity-check coded modulation,'' \emph{IEEE Transactions on Communications}, vol.~63, no.~12, pp. 4651--4665, 2015.

\bibitem{Schulte2016}
P.~Schulte and G.~Böcherer, ``Constant composition distribution matching,'' \emph{IEEE Transactions on Information Theory}, vol.~62, no.~1, pp. 430--434, 2016.

\bibitem{Amari2019}
A.~Amari, S.~Goossens, Y.~C. Gültekin, O.~Vassilieva, I.~Kim, T.~Ikeuchi, C.~M. Okonkwo, F.~M.~J. Willems, and A.~Alvarado, ``Introducing enumerative sphere shaping for optical communication systems with short blocklengths,'' \emph{Journal of Lightwave Technology}, vol.~37, no.~23, pp. 5926--5936, 2019.

\bibitem{Yoshida2020}
T.~Yoshida, M.~Karlsson, and E.~Agrell, ``Multilevel coding with flexible probabilistic shaping for rate-adaptive and low-power optical communications,'' in \emph{2020 Optical Fiber Communications Conference and Exhibition (OFC)}, 2020, pp. 1--3.

\bibitem{Sugitani2021}
K.~Sugitani, Y.~Koganei, H.~Irie, and H.~Nakashima, ``Performance evaluation of {WDM} channel transmission for probabilistic shaping with partial multilevel coding,'' \emph{Journal of Lightwave Technology}, vol.~39, no.~9, pp. 2873--2879, 2021.

\bibitem{Matsumine2022}
T.~Matsumine, M.~P. Yankov, T.~Mehmood, and S.~Forchhammer, ``Rate-adaptive concatenated multi-level coding with novel probabilistic amplitude shaping,'' \emph{IEEE Transactions on Communications}, vol.~70, no.~5, pp. 2977--2991, 2022.

\bibitem{Mirani2021}
A.~Mirani, E.~Agrell, and M.~Karlsson, ``Low-complexity geometric shaping,'' \emph{Journal of Lightwave Technology}, vol.~39, no.~2, pp. 363--371, 2021.

\bibitem{Li2021ISIT}
S.~Li, A.~Mirani, M.~Karlsson, and E.~Agrell, ``Designing {Voronoi} constellations to minimize bit error rate,'' in \emph{2021 IEEE International Symposium on Information Theory (ISIT)}, 2021, pp. 1017--1022.

\bibitem{li2023arXiv}
S.~Li, A.~Mirani, and M.~Karlsson, ``Coded modulation schemes for {Voronoi} constellations,'' \emph{arXiv preprint arXiv:2308.00407}, 2023.

\bibitem{HeZonglong2024}
Z.~He, S.~Li, E.~Deriushkina, P.~Andrekson, E.~Agrell, M.~Karlsson, and J.~Schröder, ``12.2 bit/s/{Hz} {C}-band transmission with high-gain low-complexity 24-dimensional geometric shaping,'' \emph{Journal of Lightwave Technology}, vol.~42, no.~14, pp. 4829--4836, 2024.

\bibitem{Zhao:24}
C.~Zhao, B.~Chen, Y.~Lei, S.~Li, J.~Cai, D.~Hu, W.~Fang, and L.~Sun, ``Extending the reach of multi-core fiber via {Voronoi} constellations with concatenated multilevel coding,'' \emph{Opt. Lett.}, vol.~49, no.~8, pp. 1997--2000, Apr 2024.

\bibitem{Forney1989}
G.~Forney, ``Multidimensional constellations. ii. {Voronoi constellations},'' \emph{IEEE Journal on Selected Areas in Communications}, vol.~7, no.~6, pp. 941--958, 1989.

\bibitem{Conway}
J.~Conway and N.~Sloane, ``On the {Voronoi} regions of certain lattices,'' \emph{Siam Journal on Algebraic and Discrete Methods}, vol.~5, 09 1984.

\bibitem{smith2011staircase}
B.~P. Smith, A.~Farhood, A.~Hunt, F.~R. Kschischang, and J.~Lodge, ``Staircase codes: {FEC} for 100 {Gb}/s {OTN},'' \emph{Journal of Lightwave Technology}, vol.~30, no.~1, pp. 110--117, 2011.

\end{thebibliography}

\end{document}